\documentclass[slac_one]{revtex4}
\usepackage{graphicx}
\usepackage{fancyhdr}
\usepackage{amsmath}
\usepackage{amssymb}
\newcommand{\abs}[1]{\left\lvert #1\right\rvert}

\newcommand{\vect}[1]{\mathbf{#1}}
\newcommand{\bra}[1]{\left\langle #1\right\rvert}
\newcommand{\ket}[1]{\left\lvert #1\right\rangle}
\newcommand{\Lqcd}{\Lambda_{\text{QCD}}}

\DeclareMathOperator{\Tr}{Tr}

\begin{document}

\title{Nonperturbative Effects in Event Shapes from Soft-Collinear Effective Theory}

%

\author{Christopher Lee\footnote{Electronic address: clee@phys.washington.edu}}
\affiliation{Institute for Nuclear Theory, University of Washington, Box 351550, Seattle, WA  98195-1550, USA}

\begin{abstract}
Soft-collinear effective theory (SCET) is used to demonstrate factorization for event shape distributions in the two-jet region. The leading nonperturbative power corrections to these distributions can be characterized as shape functions defined in terms of Wilson lines of soft gluons. The relation of these results to the well-known predictions for shifts in the distributions induced by the leading power corrections is discussed.
\end{abstract}

\maketitle

\thispagestyle{fancy}


\section{INTRODUCTION}
\label{sec:intro}

Event shapes provide a rich ground for testing the perturbative and nonperturbative dynamics of QCD. As they are in general less inclusive than total cross-sections, event shapes receive larger nonperturbative contributions to their observed values, but are still sufficiently inclusive to allow for a controlled expansion in these power corrections. Thus we may hope both to observe these effects easily in experiment and also to describe them theoretically.

Arguments based on analysis of renormalon ambiguities and resummed perturbation theory have led to well-known universality relations amongst the leading nonperturbative power corrections to traditional event shapes such as thrust and $C$-parameter (see \cite{review1,review2} for reviews). Similar analyses have shown the power corrections to recently-introduced angularity variables \cite{BKS} also to obey a very simple universal scaling rule \cite{scaling1,scaling2}. These arguments, however, rely  on the approximation of single gluon emission dominating the nonperturbative effects.

Effective field theory provides another framework to investigate the effects of perturbative and nonperturbative physics in hadronic observables in an organized way. By separating the physics occurring at disparate time or energy scales in a systematic expansion in a small parameter determined by the ratio of these scales, effective field theories can facilitate the proofs of factorization theorems and make manifest any relations between the nonperturbative functions or parameters contributing to different physical observables.

The soft-collinear effective theory (SCET) was introduced at first to study the energetic and light hadronic products in heavy meson decays in endpoint regions of phase space, for example, in $B\rightarrow X_s\gamma$ or $B\rightarrow D\pi$ \cite{SCET1,SCET2}. Recently, the same formalism was also applied to the case of jet production in $e^+ e^-$-annihilation or $Z$ decays. Ref.~\cite{BMW} proved a factorization theorem for jet energy distributions, which was extended to other event shapes in Ref.~\cite{BLMW} . These works reproduced in the effective theory language the factorization theorems already derived in the perturbative QCD analyses of Korchemsky and Sterman using the eikonal approximation \cite{KS95,KS99}.  In these approaches, information about the nonperturbative dynamics is contained in shape functions of Wilson lines of soft gluons. By comparing the shape functions arising for the different event shape distributions, we can learn to what extent the nonperturbative effects in the various observables are really related to each other.

In this presentation, we derive shape functions for thrust, jet mass, $C$-parameter, and angularity distributions and explore how to reproduce the relations that exist between the power corrections to these event shapes in the approach introduced by Dokshitzer and Webber \cite{DW1,DMW,DW2}. The relation between thrust and the jet mass sum becomes immediately apparent, while the others can be reproduced if we also make the single gluon approximation. In the companion contribution \cite{A001} in these Proceedings, we observe that the shape functions derived here already contain enough information to reproduce the same relations for these variables even without invoking the single gluon approximation.

\section{Event Shapes and Power Corrections}
\label{sec:events}

Some common event shapes used to describe the final state $N$ in $e^+ e^-$ annihilation are the thrust:
\begin{equation}
T = \max_{\vect{\hat t}}\frac{\sum_{i\in N}\abs{ \vect{p}_i\cdot\vect{\hat t}}}{\sum_{i\in N}\abs{\vect{p}_i}},
\end{equation}
the jet mass sum:
\begin{equation}
\widehat M^2 = \frac{1}{Q^2}(M_A^2 + M_B^2),
\end{equation}
where $M_a^2$ and $M_b^2$ are the total invariant masses of the particles in the two hemispheres $A,B$ determined by the thrust axis,
and the $C$-parameter:
\begin{equation}
C = 3(\lambda_1\lambda_2 + \lambda_2\lambda_3 + \lambda_3\lambda_1),
\end{equation}
where $\lambda_i$ are the eigenvalues of the $3\times 3$ matrix:
\begin{equation}
\theta^{rs} = \frac{1}{Q}\sum_{i\in N}\frac{\vect{p}_i^r\vect{p}_i^s}{\abs{\vect{p}_i}}.
\end{equation}
The $C$-parameter can also be expressed as:
\begin{equation}
C = \frac{3}{2}\frac{\sum_{i,j}\abs{\vect{p}_i}\abs{\vect{p}_j}\sin^2\theta_{ij}}{\left(\sum_i\abs{\vect{p}_i}\right)^2}.
\end{equation}
In addition, the angularities are defined as:
\begin{equation}
\tau_a = \frac{1}{Q}\sum_{i\in N} E_i (\sin\theta_i)^a(1-\abs{\cos\theta_i})^{1-a},
\end{equation}
which are infrared-safe variables for $-\infty<a<2$ (although we will consider only $a<1$), and where $\theta_i$ measures the angle between $\vect{p}_i$ and  the thrust axis $\vect{\hat t}$. For $a= 0$, this reduces to $\tau = 1-T$. The two-jet limit corresponds to $\tau_a,C,\widehat M^2\rightarrow 0$.

For these and other event shapes $e$, it has been argued that nonperturbative effects simply shift the argument of perturbatively-calculated distributions, that is,
\begin{equation}
\label{shift}
\frac{d\sigma}{d e}(e)\biggr\rvert_{\text{PT}} \underset{\text{NP}}{\longrightarrow} \frac{d\sigma}{de}(e - c_e \mathcal{A})\biggr\rvert_{\text{PT}} ,
\end{equation}
or just the mean values,
\begin{equation}
\langle e\rangle_{\text{PT}} \underset{\text{NP}}{\longrightarrow} \langle e\rangle_{\text{PT}} + c_e\mathcal{A},
\end{equation}
where $\mathcal{A}$ is a universal quantity whose size is of order $\mathcal{O}(\Lqcd/Q)$, and $c_e$ is an observable-dependent coefficient. For $\tau_a$, $C$, and $M^2$,
\begin{equation}
\label{Cfactors}
c_{\tau_a} = \frac{2}{1-a},\qquad c_{C} = 3\pi,\qquad c_{\widehat M^2} = 2.
\end{equation}
The quantity $\mathcal{A}$ depends on the strong coupling constant $\alpha_s(M_Z)$ and an ``effective'' coupling $\alpha_0(\mu_I)$ at an infrared scale $\mu_I$. The universality of $\mathcal{A}$ can be tested by fitting the couplings $\alpha_0$ and $\alpha_s$ to data for different event shapes (see, e.g., \cite{review1,review2}). 

\section{Factorization in SCET}
\label{sec:SCET}

To separate perturbative and nonperturbative contributions to event shape distributions, we must show that the theoretical expressions for these distributions factorize. We use the formalism of SCET to accomplish this.

The differential cross section in the variable $e$ in $e^+e^-$ annihilation to jets is given in QCD by the expression:
\begin{equation}
\label{dsigma}
\frac{d\sigma}{de} = \frac{1}{2Q^2}\sum_{N}\abs{\bra{N} J_{\text{QCD}}^\mu(0)\ket{0}L_\mu}^2 (2\pi)^4\delta^4(P - p_N)\delta(e - e(N)),
\end{equation}
where $L_\mu$ is the leptonic part of the scattering amplitude, and the current generating the final hadronic state is
\begin{equation}
\label{QCDcurrent}
J^\mu_{\text{QCD}} = \bar q\Gamma^\mu q
\end{equation}
where $\Gamma^\mu$ is the Dirac structure coupling to photons or $Z$ bosons. We suppress an implicit sum over flavors and colors.

SCET is an effective theory of QCD containing collinear and ultrasoft quarks and gluons\footnote{We are working in the theory usually known as $\text{SCET}_{\rm I}$ \cite{SCET2}.}. These degrees of freedom may be characterized by the typical size of the components of their momenta in light-cone coordinates. In terms of light-cone directions $n,\bar n$ (for instance, $n = (1,0,0,1), \bar n = (1,0,0,-1)$), these components are $p = (p^+,p^-,p^\perp)$, where $p^+ = n\cdot p$ and $p^- = \bar n\cdot p$. Collinear fields in the $n$ direction represent those particles with momenta scaling as:
\begin{equation}
p_c \sim Q(\lambda^2, 1,\lambda),
\end{equation}
while ultrasoft particles have momenta scaling as:
\begin{equation}
p_{us} \sim Q(\lambda^2,\lambda^2,\lambda^2).
\end{equation}
Here, $\lambda$ is a small parameter determined by the different energy scales appearing in the physical process at hand. In the case of $e^+e^-$ annihilation to jets, $\lambda = \sqrt{\Lqcd/Q}$, so that collinear particles have typical virtualities of order $\sim\sqrt{Q\Lqcd}$, while ultrasoft particles have virtualities $\sim\Lqcd$. The effective theory is then formulated order-by-order by expanding the QCD Lagrangian in powers of $\lambda$. \cite{SCET2}

At leading order in $\lambda$, the QCD current $J^\mu_{\text{QCD}}$ matches onto a current in SCET \cite{BMW,BLMW}:
\begin{equation}
\label{SCETcurrent}
J^\mu_{\text{SCET}} = [\bar\xi_n W_n]\Gamma^\mu_\perp C(\mathcal{\bar P},\mathcal{\bar P^\dag},\mu)[W_{\bar n}^\dag \xi_{\bar n}],
\end{equation}
where $\xi_{n,\bar n}$ are collinear fields in the back-to-back light-cone directions $n,\bar n$, $W_{n,\bar n}$ are Wilson lines of collinear gluon fields $A_n$:
\begin{equation}
W_n(z) = P\exp\left[ig\int_0^\infty ds\,\bar n\cdot A_n(\bar n s+z)\right],
\end{equation}
which are required to ensure collinear gauge invariance \cite{SCETgauge},  $C(\mathcal{\bar P},\mathcal{\bar P^\dag},\mu)$ is the Wilson coefficient, and $\mathcal{P},\mathcal{\bar P^\dag}$ are label operators picking out the large label momenta of the collinear fields \cite{SCETlabels}. At tree level in matching, $C = 1$.

Factorization is much more easily proven after the field redefinition \cite{SCETgauge}:
\begin{equation}
\label{BPS}
\xi_n\rightarrow Y_n^\dag\xi_n^{(0)},\qquad A_n\rightarrow Y_n^\dag A_n^{(0)} Y_n,
\end{equation}
using the Wilson line of ultrasoft gluons $A_{us}$:
\begin{equation}
Y_n(z) = P\exp\left[ig\int_0^\infty ds\,n\cdot A_{us}(ns+z)\right].
\end{equation}
This field redefinition removes all couplings between collinear partons and ultrasoft gluons in the leading-order SCET Lagrangian.  For instance, the only coupling between collinear quarks and ultrasoft gluons in this Lagrangian is:
\begin{equation}
\mathcal{L}_{\xi A_{us}} = \bar\xi_n in\cdot D_{us} \xi_n \longrightarrow \bar\xi_n^{(0)} in\cdot\partial \xi_n^{(0)},
\end{equation}
where the ultrasoft covariant derivative $D_{us}^\mu = \partial_{us}^\mu - igA_{us}^\mu$ turns into an ordinary derivative upon the field redefinition in Eq.~(\ref{BPS}). This eliminates interactions between the collinear and ultrasoft sectors of SCET at leading order in $\lambda$. However, the QCD current (\ref{QCDcurrent}) now matches onto a redefined SCET current:
\begin{equation}
J_{\text{SCET}}^\mu = [\bar\xi_n W_n Y_n]\Gamma^\mu_\perp C(\mathcal{\bar P},\mathcal{\bar P^\dag},\mu)[Y_{\bar n}^\dag W_{\bar n}^\dag \xi_{\bar n}],
\end{equation}
where we now suppress the superscripts $^{(0)}$ for convenience.
In the formula for the cross-section (\ref{dsigma}) we also split up the final state $N$ into the two collinear jets $J_1,J_2$ and ultrasoft particles $X_u$, so that, in SCET:
\begin{equation}
\frac{d\sigma}{de} = \frac{1}{2Q^2}\sum_{J_1,J_2,X_u}\abs{\bra{J_1 J_2 X_u}T[\bar\xi_n W_n Y_n]\Gamma^\mu_\perp C(\mathcal{\bar P},\mathcal{\bar P^\dag},\mu)[Y_{\bar n}^\dag W_{\bar n}^\dag \xi_{\bar n}]\ket{0}L_\mu}^2 (2\pi)^4(P - p_{J_1} - p_{J_2} - k_{X_u}) \delta(e - e(N)).
\end{equation}
To focus only on the leading nonperturbative effects, we work to leading order in perturbation theory, so that $J_{1,2}$ are composed of a single collinear quark or antiquark, and $C = 1$. Since there are no interactions between the collinear and ultrasoft sectors of the theory at leading order in $\lambda$, the matrix element factorizes, and we have:
\begin{equation}
\begin{split}
\frac{d\sigma}{de} &= \frac{1}{2Q^2}\int d\Pi_2 \abs{\bra{q_n\bar q_{\bar n}}[\bar\xi_n W_n]\Gamma_\perp^\mu[W_{\bar n}^\dag\xi_{\bar n}]\ket{0}L_\mu}^2\frac{1}{N_C}\Tr\sum_{X_u}\abs{\bra{X_u}T[Y_n Y_{\bar n}^\dag]\ket{0}}^2 \\
&\quad \times(2\pi)^4(P - p_q - p_{\bar q} - k_{X_u}) \delta(e - e(q,\bar q,X_u)),
\end{split}
\end{equation}
where the trace is over colors. Performing the integrals with the delta functions, the collinear part of the formula gives the leading-order total cross-section $\sigma_0$ in perturbation theory, leaving a nonperturbative shape function:
\begin{equation}
\label{shapefunction}
\frac{1}{\sigma_0}\frac{d\sigma}{de} = \frac{1}{N_C}\Tr\sum_{X_u}\abs{\bra{X_u}T[Y_n Y_{\bar n}^\dag]\ket{0}}^2\delta (e - e(X_u)) \equiv S_e(e),
\end{equation}
where we have used that for an event shape $e$ which approaches zero in the two-jet limit, the collinear partons do not contribute to $e$ at linear order in an expansion in $k_u/Q$. From now on, we will keep the dependence of the event shapes on the ultrasoft momenta only to linear order in $k_u/Q$. By studying this shape function for different choices of $e$, we may hope to uncover some useful relations between the nonperturbative effects in different event shapes.

\section{Nonperturbative Power Corrections}
\label{sec:power}

The nonperturbative shape functions for different event shapes are distinguished solely by the final delta function appearing in Eq.~(\ref{shapefunction}). For the angularities, this delta function becomes
\begin{align}
\label{angularitydelta}
\delta\left(\tau_a - \frac{1}{Q}\left(\sum_{\alpha\in A}\abs{\vect{k}_{u\perp}^\alpha}^a(k_u^-)^{1-a} + \sum_{\alpha\in B}\abs{\vect{k}_{u\perp}^\alpha}^a(k_u^+)^{1-a}\right)\right),
\end{align}
where $A,B$ are the hemispheres (determined by the thrust axis) containing the quark or antiquark, respectively, and sums are over particles $\alpha$ in the state $X_u$. For the $C$-parameter, the delta function is
\begin{equation}
\delta\left(C - \frac{1}{Q}\sum_{\alpha\in X_u}\frac{3\abs{\vect{k}_\perp^\alpha}^2}{\abs{\vect{k}^\alpha}}\right).
\end{equation}
For the jet mass sum, we have
\begin{equation}
\delta\left(\widehat M^2 - \frac{1}{Q}\left(k_u^{(A)-} + k_u^{(B)+}\right)\right),
\end{equation}
where $k_u^{(A,B)}$ is the total momentum of the ultrasoft particles in hemisphere $A,B$. From this formula and Eq.~(\ref{angularitydelta}) for $a=0$, it follows that the first moment of the thrust ($\tau_0$) and jet mass sum shape functions are identical, reproducing the result in Eq.~(\ref{Cfactors}) that $c_{\tau_0} = c_{\widehat M^2}$.
However, it is not immediately apparent from these results how the shape functions given by Eq.~(\ref{shapefunction}) for general angularities and the $C$-parameter satisfy any simple universality relations.

It is straightforward, however, to recover the relations given by Eqs.~(\ref{shift},\ref{Cfactors}) in the usual single-gluon approximation. Restricting to final ultrasoft states $X_u$ containing the single gluon $g(k)$,
\begin{equation}
\label{shape1gluon}
S_e(e) = \frac{1}{N_C}\Tr\int\frac{d^3\vect{k}}{(2\pi)^3}\abs{\bra{g(k)}T[Y_n Y_{\bar n}^\dag]\ket{0}}^2\delta(e - e(g))\delta(\abs{\vect{k}_\perp} - \abs{\vect{k}}\sin\theta),
\end{equation}
where we integrate over angles at some fixed value of the transverse gluon momentum (cf. \cite{CataniWebber}), which sets the scale for the couplings $\alpha_0,\alpha_s$. We may imagine integrating the distributions $d\sigma/de$ over a region of $e$ of size $\Delta$, near the two-jet endpoint $e\sim 0$, taking the size of $\Delta$ to be $\Lqcd/Q \ll \Delta \ll 1$. Alternatively, we may take moments of the distributions, which are dominated by events in the region of size $\Delta$ near the endpoint. We may then Taylor expand the delta functions:
\begin{align}
\delta(\tau_a - \tau_a(g(k))) &\longrightarrow \delta(\tau_a) - \frac{1}{Q}\delta'(\tau_a)\abs{\vect{k}_\perp}^a(k^{-,+})^{1-a} + \cdots, \\
\delta(C - C(g(k))) &\longrightarrow 
\delta(C) - \frac{3}{Q}\delta'(C)\frac{\abs{\vect{k}_\perp}^2}{\abs{\vect{k}}} + \cdots,
\end{align}
where in the first line, we take $k^-$ ($k^+$) if the gluon is in the hemisphere with the quark (antiquark).
The first terms of these expansions give the leading-order perturbative result for the shape functions, while the second term gives rise to the leading nonperturbative correction. Inserting these expansions back into Eq.~(\ref{shape1gluon}), evaluating the amplitude to create one gluon, and evaluating the integral, we find the relation between the leading nonperturbative shift to the first moments of the angularity and $C$-parameter distributions:
\begin{equation}
\delta\langle \tau_a\rangle = \frac{2}{1-a}\frac{1}{3\pi}\delta\langle C\rangle,
\end{equation}
which are consistent with the values for the factors $c_e$ given in Eq.~(\ref{Cfactors}).

In \cite{A001}, it is shown how to derive the universality relations directly from the shape functions (\ref{shapefunction}) without resorting to the single gluon approximation.

\section{Conclusions}

We have recast the factorization of event shape distributions in the language of soft-collinear effective theory. The decoupling of collinear and ultrasoft modes at leading-order in the SCET expansion allows for a simple proof of this factorization. The leading nonperturbative effects in the event shape distribution for $e$ are contained in the shape function $S_e(e)$, encoding the effect of ultrasoft gluon emission in the form of Wilson lines. Here we have seen how to reproduce the usual relations for the leading nonperturbative effects shifting the first moments of the thrust and jet mass sum, and, if we make the single gluon approximation, also of the angularity and $C$-parameter distributions.

\begin{acknowledgments}
It is a pleasure to thank the high energy theory groups at UCSD and Caltech for their generous hospitality during portions of this work, the nuclear theory group at Caltech for its support during its completion, and M. Wise, A. Manohar, and C. Bauer for their discussions and collaboration. This work was supported in part by the U.S. Department of Energy under grant number DE-FG02-00ER41132.

\end{acknowledgments}



\begin{thebibliography}{99} 

\bibitem{review1}
  M.~Beneke and V.~M.~Braun,
  arXiv:hep-ph/0010208.
  
\bibitem{review2}
  M.~Dasgupta and G.~P.~Salam,
  J.\ Phys.\ G {\bf 30}, R143 (2004)
  [arXiv:hep-ph/0312283].


\bibitem{BKS}
  C.~F.~Berger, T.~Kucs and G.~Sterman,
  Phys.\ Rev.\ D {\bf 68}, 014012 (2003)
  [arXiv:hep-ph/0303051].
  
\bibitem{scaling1}
  C.~F.~Berger and G.~Sterman,
  JHEP {\bf 0309}, 058 (2003)
  [arXiv:hep-ph/0307394].
  
\bibitem{scaling2}
  C.~F.~Berger and L.~Magnea,
  Phys.\ Rev.\ D {\bf 70}, 094010 (2004)
  [arXiv:hep-ph/0407024].
  
\bibitem{SCET1}
  C.~W.~Bauer, S.~Fleming and M.~E.~Luke,
  Phys.\ Rev.\ D {\bf 63}, 014006 (2001)
  [arXiv:hep-ph/0005275].
  
\bibitem{SCET2}
  C.~W.~Bauer, S.~Fleming, D.~Pirjol and I.~W.~Stewart,
  Phys.\ Rev.\ D {\bf 63}, 114020 (2001)
  [arXiv:hep-ph/0011336].
  
\bibitem{BMW}
  C.~W.~Bauer, A.~V.~Manohar and M.~B.~Wise,
  Phys.\ Rev.\ Lett.\  {\bf 91}, 122001 (2003)
  [arXiv:hep-ph/0212255].
  
\bibitem{BLMW}
  C.~W.~Bauer, C.~Lee, A.~V.~Manohar and M.~B.~Wise,
  Phys.\ Rev.\ D {\bf 70}, 034014 (2004)
  [arXiv:hep-ph/0309278].
  
\bibitem{KS95}
  G.~P.~Korchemsky and G.~Sterman,
  Nucl.\ Phys.\ B {\bf 437}, 415 (1995)
  [arXiv:hep-ph/9411211].
  
\bibitem{KS99}
  G.~P.~Korchemsky and G.~Sterman,
  Nucl.\ Phys.\ B {\bf 555}, 335 (1999)
  [arXiv:hep-ph/9902341].
  
\bibitem{DW1}
  Y.~L.~Dokshitzer and B.~R.~Webber,
  Phys.\ Lett.\ B {\bf 352}, 451 (1995)
  [arXiv:hep-ph/9504219].
  
\bibitem{DMW}
  Y.~L.~Dokshitzer, G.~Marchesini and B.~R.~Webber,
  Nucl.\ Phys.\ B {\bf 469}, 93 (1996)
  [arXiv:hep-ph/9512336].
  
\bibitem{DW2}
  Y.~L.~Dokshitzer and B.~R.~Webber,
  Phys.\ Lett.\ B {\bf 404}, 321 (1997)
  [arXiv:hep-ph/9704298].


\bibitem{A001}
C.~Lee and G.~Sterman, ``Universality of Nonperturbative Effects in Event Shapes,'' in these Proceedings [arXiv:hep-ph/0603066].  

   
\bibitem{SCETgauge}
  C.~W.~Bauer, D.~Pirjol and I.~W.~Stewart,
  Phys.\ Rev.\ D {\bf 65}, 054022 (2002)
  [arXiv:hep-ph/0109045].
  
   
\bibitem{SCETlabels}
  C.~W.~Bauer and I.~W.~Stewart,
  Phys.\ Lett.\ B {\bf 516}, 134 (2001)
  [arXiv:hep-ph/0107001].
  
\bibitem{CataniWebber}
  S.~Catani and B.~R.~Webber,
  Phys.\ Lett.\ B {\bf 427}, 377 (1998)
  [arXiv:hep-ph/9801350].

  
\end{thebibliography}
\end{document}